\documentclass[twocolumn,showpacs,aps,prl,amsmath,amssymb,floatfix]{revtex4}
\usepackage{graphicx}
\usepackage{dcolumn}
\usepackage{bm}
\begin{document}
\title{Spin-orbit interactions mediated negative differential resistance in a
quasi-two-dimensional electron gas with finite thickness}
\author{E. Nakhmedov$^{1,2}$, O. Alekperov$^{2}$, and R. Oppermann$^{1}$}
\affiliation{ $^1$Institut f\"ur Theoretische Physik,
Universit\"at W\"urzburg, Am Hubland,
D-97074 W\"urzburg, Germany\\
$^2$Institute of Physics, Azerbaijan National Academy of
Sciences,H. Cavid str. 33, AZ1143 Baku, Azerbaijan}
\date{\today}
\begin{abstract}
{Effects of the spin-orbit interactions on the energy spectrum,
Fermi surface and spin dynamics are studied in structural- and
bulk-inversion asymmetric quasi-two-dimensional structures with a
finite thickness in the presence of a parabolic transverse
confining potential. One-particle quantum mechanical problem in
the presence of an in-plane magnetic field is solved numerically
exact. Interplay of the spin-orbit interactions, orbital- and
Zeeman-effects of the in-plane magnetic field yields a
multi-valley subband structure, typical for realization of the
Gunn effect. A possible Gunn-effect-mediated spin accumulation is
discussed.}
\end{abstract}
\maketitle
One of the major goals of modern electronics is the search for new
spin-involved functionalities \cite{als02} in solid state
nanoscale devices. Electric field controlled Rashba spin-orbit
(SO) coupling \cite{rashba60}, which is originated from large
potential gradient on the semiconductor/insulator interface of
quantum wells and MOSFETs in the presence of macroscopic
structural inversion asymmetry (SIA), is a promising tool
\cite{dd90} in a realization of spin transport devices. On the
other hand, special semiconduction materials with a bulk inversion
asymmetry (BIA) in their crystalline structure produce so-called
Dresselhaus SO interaction \cite{dress55}, which interplays with
Rashba interaction yielding unusual physical effects \cite{cl08,dls10}.

Different experimental techniques have been developed recently to control 
a coupling of spin to the electric field \cite{kmdg03,kmga04,lbr07}. 
An efficient $\hat{g}$-tensor modulation resonance, observed in a 
parabolic $Al_xGa_{1-x}As$ quantum well \cite{kmdg03} with varying 
$Al$ content $x=x(z)$ across the well, provided an opportunity to manipulate electron 
spins by means of various electron spin resonance type techniques.
In-plane magnetic field in all of these experiments seems to be 
rather favorable to get a pronounced spin resonance. On the other hand, 
gate voltage control of spin dynamics in practical devices was shown \cite{re03} 
to be reasonable only for a finite thickness of the electron gas.  
SO interactions in a 2D electronic system produce an effective 
in-plane field, which results in an drift-driven in-plane spin 
polarization \cite{edelstein90}. External in-plane magnetic field 
appears to be not always algebraically summed with SO induced 
effective field and results in a surprizing out-of-plane spin 
polarization \cite{erh07}, which was observed in strained
$n-InGaAs$ film \cite{kmga04}. On the other hand Hanle precession of optically
oriented 2D electrons in $GaAs$ \cite{kk90} can be well described by 
algebraic addition of these fields. All these facts show nontrivial effects of
in-plane magnetic field on spin dynamics in quasi-2D systems.
 
In this paper we show that an interplay of Rashba and Dresselhaus
SO interactions in a quasi-two-dimensional (quasi-2D) electron gas
with finite thickness results in a multi-valley ('N'-shaped)
energy dispersion in the presence of an in-plane magnetic field.
The obtained energy-momentum relation looks like those existing in
GaAs and InP-type materials of the transferred-electron devices,
where negative differential resistance and Gunn microwave current
oscillations take place. Finite thickness of the electron gas
gives rise to transverse-quantized levels with Rashba-Dresselhaus
spin-split subbands in each level. The in-plane magnetic field
removes degeneration in the anticrossing points of different
subbands, opening a magnetic field-controlled gap at these points.
Barrier height between the valleys in the 'N'-shaped
energy-momentum relation and the valley curvatures seem to be
controlled by Rashba- and Dresselhaus SO coupling constants, the
in-plane magnetic field strength and Land\'e factor. The
dependence of the energy spectrum on the magnetic field shows
strongly non-linear and non-monotonic behavior even in weak field
strengths.

The single particle Hamiltonian of the system in the effective
mass approximation is written as
 \begin{equation}
\hat{H} = \frac{{\bf P}^2}{2 m^{\ast}} + \frac{m^{\ast} \omega_0^2
z^2}{2} - e E_g z + \hat{H}_{so} + \frac{1}{2}g \mu_B {\bf
\sigma}{\bf B}, 
\label{hamiltonian}
\end{equation}
where ${\bf P}= {\bf p} - \frac{e}{c}{\bf A}$ is an electron
momentum in the presence of a vector-potential ${\bf A}$, $m^{\ast}$
and $e$ are the electronic effective mass and charge,
respectively; $E_g$ is a strength of the gate electric field. A
parabolic potential with a frequency $\omega_0$, characterizing
the electron gas thickness, does not produce SO interactions, and
is chosen in Eq.(\ref{hamiltonian}) to confine the electron gas in
$z$-direction. Since Rashba SO interaction in the conduction band
of a semiconductor is determined by the electric field in the
valence band rather than by that in the conduction band
\cite{lassnig85}, the parabolic confinement approximation neglects
a small interface contribution to Rashba SO coupling constant. The
last term in Eq.(\ref{hamiltonian}) is Zeeman splitting energy in
the external magnetic field ${\bf B}$ with $\hbar \omega_z = g
\mu_B B/2$, where $\mu_B = \frac{e \hbar}{2 m_0}$ is the Bohr
magneton of a free electron with mass $m_0$, $g$ is the effective
Land{\'e} factor, and
$\mathbf{\sigma}=\{\sigma_x,\sigma_y,\sigma_z\}$ are the Pauli
spin matrices. $\hat{H}_{so}$ in
Eq.(\ref{hamiltonian}) contains Rashba term as well as 
Dresselhaus term with characteristic parameters $\alpha$ and
$\beta$, correspondingly
\begin{equation}
\hat{H}_{so}=\frac{\alpha}{\hbar} (\sigma_x P_y -
\sigma_y P_x) + \frac{\beta}{\hbar} (\sigma_x P_x - \sigma_y P_y).
\label{spin-orbit}
\end{equation}
The electronic wave function $\Psi (x,y,z)$ can be
expressed in the form $\Psi (x,y,z) = e^{ik_x x + ik_y y}
{\psi_{\uparrow}(z) \choose \psi_{\downarrow}(z)}$ for a magnetic
field aligned along $x$-axes ${\bf B}=\{B, 0, 0\}$ under the gauge
${\bf A}=\{0, -Bz, 0\}$.
In the absence of the SO interactions and Zeeman term, Hamiltonian
becomes diagonal and the equations for $\psi_{\uparrow}$ and
$\psi_{\downarrow}$ are reduced to the oscillator equation
$\mathcal{\hat H}_0 \psi_n^{(0)}(z)= E_n \psi_n^{(0)}$ with real
wave function $\psi_n^{(0)}(z)=(\sqrt \pi 2^n n!)^{-1/2}
\exp \left[-(z-z_0)^2/(2a_B^2) \right] H_n\left((z-z_0)/a_B\right)$ and the
energy spectrum $E_n=\hbar \omega (n+1/2)-\frac{\hbar^2 k^2}{2
m^{\ast}} + \frac{(\hbar k_y \omega_B - eE_g)^2}{2 m^{\ast}
\omega^2}$, where $z_0 = \frac{eE_g - k_y \hbar \omega_B}{m^{\ast}
\omega^2}$, $a_B=(\hbar/m^{\ast}\omega)^{1/2}$, 
$\omega = (\omega_B^2 + \omega_0^2)^{1/2}$, $\omega_B=eB/m^{\ast} c$, $k =
(k_x^2+k_y^2)^{1/2}$, and $H_n(z)$ is the Hermite polynomial.
General solutions are chosen as linear combinations of
$\psi_n^{(0)}(z)$
\begin{equation}
\psi_{\sigma}(z) =
e^{-\frac{(z-z_0)^2}{2a_B^2}}\sum_{n=0}^{\infty}\frac{a^{\sigma}_n}{\sqrt{a_B
\sqrt \pi 2^n n!}} H_n\left(\frac{z-z_0}{a_B}\right),
\label{wave}
\end{equation}
where $\sigma = \uparrow, \downarrow$.
By putting Eq.(\ref{wave}) into
Schr\"odinger equation one gets the matrix equations $\hat{\bf N_{\sigma}}
{\bf a}^{\sigma}=0$ for the vectors ${\bf a}^{\sigma} =
\{a^{\sigma}_0, a^{\sigma}_1, a^{\sigma}_2, \dots \}$. $\hat{\bf N}_{\sigma}$ 
with $\sigma = \uparrow, \downarrow$ are square
penthadiagonal matrices of infinite order with non-zero entries
$N^{\sigma}_{i,j} \ne 0$ only if $|i-j| \le 2$, and
$\hat{\bf N}_{\uparrow} = (\hat{\bf N}_{\downarrow})^{\ast}$.
The energy spectrum has to be found from the secular
equation, by equating the determinant of the matrix $\hat{\bf N}$
to zero. The infinite penthadiagonal matrix is truncated down to
the first $n$ rows and $n$ columns, and the roots of its
determinant are found by numeric methods. All parameters in our
numeric calculations, shown below with tilde, are done
dimensionless in the unit of the characteristic frequency
$\omega_0$ of the confining potential or in the length scale $l_0
= (\hbar/m^{\ast} \omega_0)^{1/2}$, which is a measure of the
electron gas thickness.

\begin{figure}
\vspace{-1.8cm}
\resizebox{.47\textwidth}{!}{%
\includegraphics[width=7.0cm]{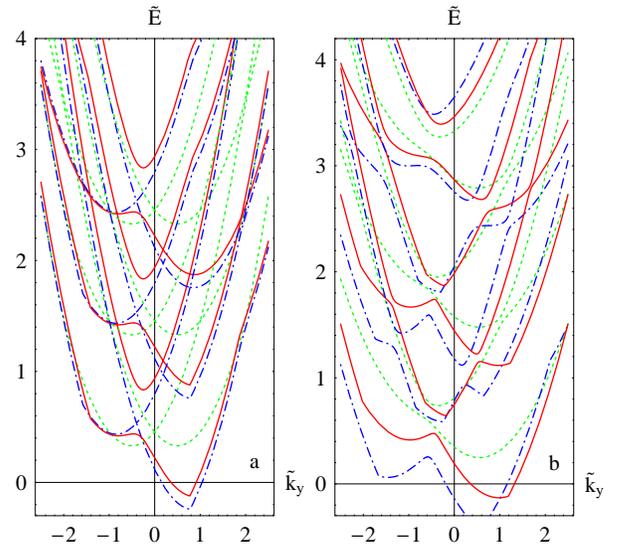}}
\vspace{-0.7cm} \caption{(Color online) Dependence of the energy spectrum on
Rashba- and Dresselhaus SO coupling constants as well as on the
orbital- and spin-effects of the in-plane magnetic field. $E$ vs.
$k_y$ is shown (a) in the absence of the magnetic field at
$\tilde{k}_x=0$, $\tilde{f}_g=0.3$, $\tilde{\alpha} = 0.5$,
$\tilde{\beta} = 0$ by dashed (green) curves; $\tilde{k}_x=0.4$,
$\tilde{f}_g=0$, $\tilde{\alpha}=\tilde{\beta} = 0.6$ by
dot-dashed (blue) curves and $\tilde{k}_x= 0.4$,
$\tilde{f}_g=0$, $\tilde{\alpha} = 0.8$, $\tilde{\beta} = 0.4$ by
solid (red) curves; (b) under the in-plane magnetic field
 $\tilde{\omega}_B = 0.7$, $\tilde{g}=0.2$ and $\tilde{k}_x = 0.2$,
$\tilde{\beta} = 0.4$ but for different values of $\tilde{\alpha}$
and $\tilde{f}_g$: $\tilde{\alpha}=0.4$, $\tilde{f}_g=0$ by dashed
(green) curves, $\tilde{\alpha} = 1.2$, $\tilde{f}_g = 0.3$ by
dot-dashed (blue), and $\tilde{\alpha} = 0.9$, $\tilde{f}_g = 0$
by solid (red) curves.} \label{dispersion} \vspace{-0.5cm}
\end{figure}

{\it Energy spectrum.} The solutions of the $n \times n$
determinant converge very rapidly as $n$ increases. 
The energy dispersion for the first three transverse quantized 
levels (n=3) is depicted in Fig.{\ref{dispersion}} for 
different values of the SO coupling constants and the magnetic field.
In the absence of the in-plane magnetic field and one of the SO
coupling constant, the energy spectrum $E_n(k_y)$ is described by
the well-known two symmetric parabolas in each
transverse-quantized level (dashed (green) curves in
Fig.\ref{dispersion}a). An external gate electric field
$\tilde{f}_g \equiv eE_gl_0/(\omega_0 \hbar)$ coherently
shifts the energy subbands. Non-zero $k_x$ component of the
momentum splits spin-up and spin-down spectra at $k_y=0$. The
energy spectrum becomes asymmetric along momentum- and energy-axes
if $\tilde{\alpha} \equiv \alpha/(\hbar \omega_0 l_0) \ne 0$, 
$\tilde{\beta} \equiv \beta/(\hbar \omega_0 l_0) \ne 0$ and $k_x \ne
0$. 
Energy spectra corresponding to the same transverse-quantized level with opposite
spin orientations split at the intersection point for $\alpha \ne
\beta$, whereas they do not split at the intersection point for
$\alpha = \beta$. Nevertheless, anticrossing of the energy spectra
corresponding to different levels persists irrespective of the
values of $\alpha$ and $\beta$. Fig.\ref{dispersion}b describes
the dependence of $E_n$ on $k_y$ in the presence of the in-plane
magnetic field $\tilde{\omega}_B \equiv \omega_B/\omega_0 $, Zeeman splitting
$\tilde{g} \equiv g m^{\ast}/(4 m_0)$ and $\tilde{k}_x \ne 0$ 
for different values of the SO coupling constants. The
magnetic field seems to remove the degeneracy at the anticrossing
points, and to open a gap, which increases with magnetic field.
Although the Zeeman splitting strongly modifies the energy levels
in the presence of only Rashba SO coupling, shifting the parabolas
bottoms along $k_y$ and splitting the energies of spin-up
and spin-down electrons, it does not change the levels symmetry in
the presence of only Dresselhaus SO coupling (dot-dashed (blue)
curves in Fig.\ref{dispersion}b). The same picture is obtained for
the energy dispersion along $k_x$ under the similar conditions
given in Fig.\ref{dispersion}b but with the replacements $k_x
\leftrightarrow k_y$ and $\alpha \leftrightarrow \beta$.
\begin{figure}
\vspace{-1.7cm}
\resizebox{.47\textwidth}{!}{%
\includegraphics[width=7cm]{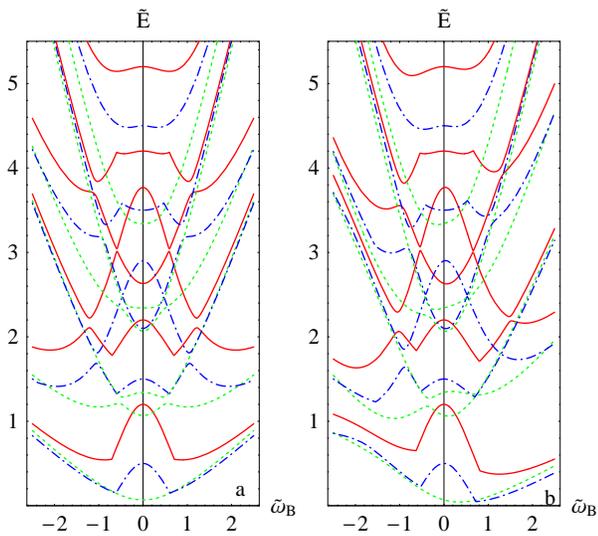}}
\vspace{-0.5cm} 
\caption{(Color online) Dependence of the energy spectrum on the
in-plane magnetic field (a) in the absence $\tilde{g}=0$ and (b)
in the presence $\tilde{g} = 0.2$ of Zeeman splitting at $\tilde{f}_g=0$. The
parameters $\tilde{k}_x =0.4$, $\tilde{k}_y = 0.5$,
$\tilde{\alpha} = 0.6$, and $\tilde{\beta} = 0.4$ correspond to
dotted (green) curve. Dashed (blue) and solid (red) curves depict
the cases $\tilde{k}_x = 0.2$, $\tilde{k}_y = 1.4$,
$\tilde{\alpha} = 0.8$, $\tilde{\beta} = 0.4$ and $\tilde{k}_x
=0.4$, $\tilde{k}_y = 1.8 $, $\tilde{\alpha} = 0.6$,
$\tilde{\beta} = 0.4$ respectively.} \label{e-mag}
\vspace{-0.5cm}
\end{figure}

The dependence of the energy spectrum on the in-plane magnetic
field under different momentum components and SO coupling
constants is given in Fig.\ref{e-mag}a for zero Land\'e factor and
in Fig.\ref{e-mag}b for a non-zero Land\'e factor,
$\tilde{g}=0.2$. The energy spectrum in both cases displays
considerably nonlinear and non-monotonic behavior even under weak
magnetic fields, when $0<\tilde{\omega}_B<1$. 
The magnetic field splits the spectrum at the
anticrossing points. The energy dispersion in the negative
magnetic field is exactly the same as the dispersion in the
positive magnetic field but with reversed in sign the Land\'e
factor.

\begin{figure}
\vspace{-1.0cm}
\resizebox{.50\textwidth}{!}{%
\includegraphics[width=7cm]{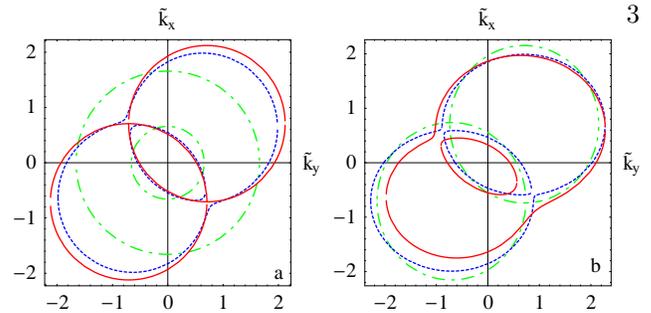}}
\vspace{-0.5cm} \caption{(Color online) Fermi surface of electrons in the first
transverse-quantized level. (a) Dot-dashed (green) curves
correspond to $\tilde{\alpha}=0.5$, $\tilde{\beta}=0$,
$\tilde{f}_g=0.3$, dashed (blue) curves to $\tilde{\alpha} = 0.5$,
$\tilde{\beta} = 0.4$, $\tilde{f}_g=0$ and solid (red) curves to
$\tilde{\alpha}=\tilde{\beta}=0.5$, $\tilde{f}_g=0$ in the absence
of magnetic field $\tilde{\omega}_B= \tilde{g} = 0$; (b)
Dot-dashed (green) curves correspond to
$\tilde{\omega}_B=\tilde{g}=0$, $\tilde{f}_g=0.3$ for
$\tilde{\alpha}=\tilde{\beta}=0.5$,  dashed (blue) curves to
$\tilde{\omega}_B = 0.7$, $\tilde{g} = 0$, $\tilde{f}_g=0$ for
$\tilde{\alpha}=\tilde{\beta}=0.5$, and solid (red) curves to
$\tilde{\omega}_B=0.7$, $\tilde{g}=0.3$, $\tilde{f}_g=0$ for
$\tilde{\alpha}=0.5$, $\tilde{\beta} = 0.4$.} 
\label{FermiSurface}
\vspace{-0.5cm}
\end{figure}

{\it Fermi surface.} Influence of different SO interactions and
in-plane magnetic field on the Fermi surface is shown in
Fig.\ref{FermiSurface} for electrons, resided the first
transverse-quantized level. The dimensionless energy is fixed in
the middle of the two adjacent transverse levels
$\tilde{E}_F \equiv E_F/(\hbar \omega_0) = 1.0$.
In the presence of only one SO interaction the Fermi surface is 
symmetrically splitted at $\omega_B=0$ into two surfaces of spin-up 
and spin-down electrons even under non-zero gate voltage. Fermi
surfaces become strongly anisotropic for $\alpha \ne 0$ and $\beta
\ne 0$ due to shifting of two circles away from each other in both
$k_x$ and $k_y$ directions. If $\alpha \ne \beta$, the circles
corresponding to spin-up and spin-down electrons split at the
intersection points. Nevertheless they do not split at the
intersection points if Rashba and Dresselhaus constants are equal
each other, $\alpha = \beta$.
Gate voltage increases the anisotropy, but it does not split the
Fermi surface at the intersection points for $\alpha = \beta$. Instead,
in-plane magnetic field removes a degeneracy at the intersection
points.

{\it Spin-Gunn effect.} Multi-valley energy dispersion, discussed
above, allow us to suggest an existence of negative differential
resistance and spin-Gunn effect \cite{gunn63} in a quasi-2D
electron gas in the presence of SO interactions. Assume that the
Fermi level crosses the second transverse-quantized level in the
energy-momentum dispersion, shown in Fig.\ref{dispersion}b with solid 
(red) curve under the conditions $\tilde{k}_x= 0.2$, $\tilde{\alpha}
= 0.9$, $\tilde{\beta} = 0.4$, $\tilde{\omega}_B \equiv B/B_0
=0.7$, and $ \tilde{g} = 0.2$. In order to estimate
$\tilde{\alpha}$, $\tilde{\beta}$, $B_0= m^{\ast}c\omega_0/e$ and
the electron gas thickness 
$l_0=\sqrt{\hbar/m^{\ast}\omega_0}$, we use $\alpha \sim 1\times
10^{-8} eV \cdot cm$, and $\beta \sim 0.4 \times
10^{-8} eV \cdot cm$, as well as $\omega_0 \hbar \sim 10~ meV$,
\cite{rashba04E}, which yield $l_0=10.6~nm$ in agreement with quantum well 
width in the experiments \cite{kmdg03,kmga04}, $B_0 = 0.9~T$ and 
$B=0.6~T$ (corresponding to $\tilde{\omega}_B=0.7$),
$\tilde{\alpha}=0.944$ and $\tilde{\beta}=0.38$ for, e.g.
conduction band electron in GaAs with $m^{\ast}=0.068 m_0$. Indeed, 
Rashba SO constant varying in the interval of 
$\alpha \sim (4.47 - 6.30)\times 10^{-9}~ eV \cdot cm$ has been reported 
\cite{czwz02} for InAlAs/InGaAs heterostructures.
Furthemore a giant SO splitting was recently
observed for quantum well states of a Bi monolayer on Ag(111)
\cite{ahem07}, on Si(111) \cite{gsf09}, and on Cu(001)
\cite{mrdw10} with Rashba parameters correspondingly $\alpha =
3.05 \times 10^{-8}~ eV \cdot cm$, $1.37 \times 10^{-8}~ eV \cdot
cm$ and $(1.5 \div 2.5) \times 10^{-8}~ eV \cdot cm$.
\begin{figure}
\vspace{-1.5cm}
\resizebox{.47\textwidth}{!}{%
\includegraphics[width=7cm]{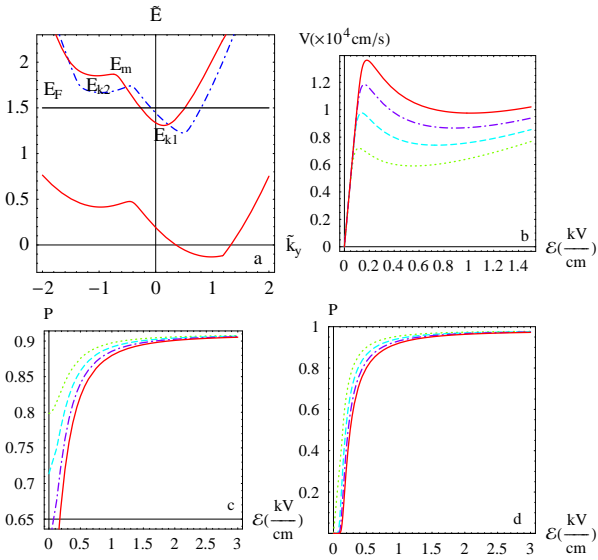}}
\vspace{-0.5cm} \caption{ (Color online) (a) Two valley spectrum corresponding to
the lower branch of the first and the second transverse-quantized
levels with $\tilde{\alpha}=0.9$, $\tilde{\beta} = 0.4$,
$\tilde{\omega}_B=0.7$, $\tilde{g}=0.2$, $\tilde{f}_g=0$ and
$\tilde{k}_x=0.2$ shown by solid (red) curves, and
$\tilde{\alpha}=1.2$, $\tilde{\beta} = 0.4$,
$\tilde{\omega}_B=0.8$, $\tilde{g}=0.3$, $\tilde{f}_g=0.3$ and
$\tilde{k}_x=0.4$ shown by dot-dashed (blue) curve. The dependence
of (b) the average drift velocity $v$ and (d) the spin
accumulation $P$ on the electric field $\mathcal{E}$ for
$\mu_1=10~m^2/V \cdot s$ and $T = 4.2K$, $\mu =\mu_2/\mu_1= 0.02$,
$R = 50$. (c) $P$ vs. $\mathcal{E}$ for $\mu_1=20~m^2/V \cdot s$
and $T=50K$, $\mu=\mu_2/\mu_1=0.1$, $R=10$. Four different values
of the activation energy $\Delta E = 4~meV,~6~meV, ~8~meV$ and
$10~meV$ correspond to dotted (green), dashed (violet), dot-dashed
(blue), and solid (red) curves in each figures.}
\label{polarization} 
\vspace{-0.5cm}
\end{figure}

In equilibrium most electrons reside near the bottom $E_{k1}$ of
the lower valley in Fig.\ref{polarization}a, which corresponds to
an anticrossing point with small effective mass $m_1^{\ast}$ and
higher mobility $\mu_1$ due to the sharp curvature of the valley.
The electric field, applied along $k_y$ axes, bends the Fermi
level and accelerates the electrons with a definite (up- or
down-) spin polarization to the top of the valley separation $E_m$
when the field reaches a threshold value. The upper valley is more
flat where the effective mass $m_2^{\ast}$ is rather heavier and
the mobility $\mu_2$ is smaller than $\mu_1$. The ratio
$m_2^{\ast}/m_1^{\ast}$ is estimated to be $\sim 80$ by fitting
the valleys' bottoms to parabolas. Our calculations show that the 
effective masse in a valley, obtained by anticrossing of a parabola
of $n$th subband with a parabola of $(n+1)$th subband, can be controlled. Indeed, by 
changing the signs or the amounts of SO coupling constants, magnetic field,
gate voltage and $g$-factor the bottoms of the parabolas are controllably 
shifted in different directions along the energy- and momentum-axces, giving different
curvatures and effective masses in the valleys. 
The total electron density $n$
is written as a sum of the densities $n_1$ and $n_2$ in the lower
and upper valleys, correspondingly, $n=n_1 + n_2$. The population
ratio between the upper and lower valleys is assumed to be given
by Maxwellian energy distribution as $n_2/n_1= R \exp (-\Delta
E/kT_e)$, where $T_e$ is the electron temperature, $R$ is the
ratio of the density of states in the upper $N_2$ and lower $N_1$
valleys with $R=m_2^{\ast}/m_1^{\ast}$ for quasi-2D electron gas,
and $\Delta E = E_{k2} - E_{k1}$ is the energy separation between
the two valley minima, which can be estimated from e.g.
Fig.\ref{polarization}a to be $\Delta E = 0.45~\hbar \omega_0$.
Then, electron accumulation $P$ with definite spin polarization in
the upper valley is written as
\begin{equation}
P = \frac{n_2}{n} = \frac{n_2}{n_1 + n_2} = \frac{R \exp(-\Delta
E/k T_e)}{1 + R \exp(-\Delta E/k T_e)}, \label{accumulation}
\end{equation}
The electron temperature $T_e$ is defined from the condition $e
\mathcal{E} v \tau_e = 3k(T_e - T)/2$, meaning that electrons gain
an energy $e \mathcal{E} v \tau_e$ from the electric field
$\mathcal{E}$, aligned in $y$-axis, as they move up the potential
barrier with the average drift velocity $v=(\mu_1 n_1 + \mu_2 n_2)
\mathcal{E}/(n_1 + n_2)$, and transfer the excess energy to the
atomic lattice with temperature $T$ in the energy relaxation time
$\tau_e$, \cite{sze07}. The steady-state current $J$ is expressed
through $v$ as $J = e (\mu_1 n_1 + \mu_2 n_2) \mathcal{E} = e n
v$, which is calculated from the following self-consistent
equations
\begin{equation}
T_e = T + \frac{2 e \tau_e v \mathcal{E}}{3k}, \quad v=\mu_1
\mathcal{E} \frac{1 + (\mu_2/\mu_1) R \exp(-\Delta E/ k T_e)}{1 +
R \exp(-\Delta E/ k T_e)}. \label{T}
\end{equation}
Eqs.(\ref{accumulation}) and (\ref{T}) determine completely
$v-\mathcal{E}$ or $I-V$ dependence, which displays a shape with a
region of negative differential resistance.
Numerical solution of Eqs.(\ref{accumulation}) and (\ref{T})
yields a dependence of the spin-polarized electron accumulation $P$ on $\mathcal{E}$,
which is presented in Fig.\ref{polarization}c, and d for temperatures 
$T = 50 K$ and $4.2 K$, correspondingly. $v$ vs. $\mathcal{E}$ 
and $P$ vs. $\mathcal{E}$ dependences are depicted in
each figures for different valley separations $\Delta E = 4~meV,
6~meV, 8~meV$ and $10~meV$, which are shown by dotted (green),
dashed (violet), dot-dashed (blue), and solid (red) curves,
respectively. High temperature mobility of electrons in, e.g. GaAs
varies in the interval of $(10 \div 20)~m^2/V \cdot s$
\cite{sze07}. 
The energy relaxation time $\tau_e$
is chosen to be $\tau_e = 10^{-12}~s$. The higher valley is
basically resided at $T= 50 K$ by electrons, according to
Fig.\ref{polarization}c, due to thermal fluctuations even in the
absence of the electric field. The spin-polarized electron accumulation slightly
increases from $P \sim 0.6$ up to the saturation value $P \sim
0.90$ as $\mathcal{E}$ increases and reaches $\sim 1.5~kV/cm$.
The spin-dependent electron accumulation at
$T=4.2K$, given in Fig.\ref{polarization}d for $\mu_1 = 10~m^2/V \cdot s$,
$\mu_2/\mu_1 = 0.02$ and $m_2^{\ast}/m_1^{\ast} = 50$, sharply
increases with electric field from $P\sim 0.0$ up
to $P\sim 0.98$ at $\mathcal{E}\sim 1~kV/cm$. Fermi energy
moves up with increasing the electron concentration and it crosses
a subband with opposite spin polarization.

In conclusion, we show that interplay of SO interactions and
in-plane magnetic field strongly changes an electron spectrum of
quasi-2D electron gas with finite thickness yielding multi-valley
energy dispersion in each transverse-quantized subband. As a
result, an instable regime of negative differential conductance is
realized for spin-polarized electrons. 


\vspace{-0.5cm}
\begin{thebibliography}{99}
\vspace{-0.5cm}
\bibitem{als02}{\it Semiconductor Spintronics and Quantum
Computation}, Eds. D. D. Awschalom et al. (Springer, Berlin, 2002).
\bibitem{rashba60} E. I. Rashba, Fiz. Tverd. Tela (Leningrad) {\bf
2}, 1224 (1960) [Sov. Phys. Solid State {\bf 2}, 1109 (1960)].
\bibitem{dd90} S. Datta and B. Das, Appl. Phys. Lett. {\bf 56},
665 (1990).
\bibitem{dress55} G. Dresselhaus, Phys. Rev. {\bf 100}, 580
(1955).
\bibitem{cl08} O. Chalaev and D. Loss, Phys. Rev. B {\bf 77},
115352 (2008).
\bibitem{dls10} M. Duckheim et al., Phys. Rev. B {\bf 81}, 085303 (2010).
\bibitem{kmdg03} Y. Kato et al., Science {\bf 299}, 1201 (2003).
\bibitem{kmga04} Y. K. Kato et al., Phys. Rev. Lett. {\bf 93}, 176601 (2004). 
\bibitem{lbr07} E. A. Larid et al., Phys. Rev. Lett. {\bf 99}, 246601 (2007).
\bibitem{re03} E. I. Rashba and Al. L. Efros, Phys. Rev. Lett. {\bf 91},
126405; Phys. Rev. B {\bf 73}, 165325 (2006).
\bibitem{edelstein90} V. M. Edelstein, Solid State Commun. {\bf 73}, 233 (1990).
\bibitem{erh07} H.-A. Engel, E. I. Rashba, and B. I. Halperin, Phys. Rev. Lett.
{\bf 98}, 036602 (2007).
\bibitem{kk90}V. Kalevich and V. Korenev, JETP Lett. {\bf 52}, 230 (1990).
\bibitem{lassnig85} R. Lassnig, Phys. Rev. B {\bf 31}, 8076 (1985).
\bibitem{gunn63} J. B. Gunn, Solid State Commun. {\bf 1}, 88
(1963).
\bibitem{rashba04E} E. I. Rashba, Physica E {\bf 20}, 189 (2004).
\bibitem{czwz02} L. J. Cui et al., Appl. Phy. Lett. {\bf 80}, 3132 (2002).
\bibitem{ahem07}C. R. Ast et al., Phys. Rev. Lett. {\bf 98}, 186807 (2007).
\bibitem{gsf09} I. Gierz et al., Phys. Rev. Lett. {\bf 103}, 046803 (2009).
\bibitem{mrdw10} S. Mathias et al., Phys. Rev. Lett. {\bf 104}, 066802 (2010).
\bibitem{sze07} S. M. Sze and Kwok K. Ng, {\it Physics of
Semiconductor Devices}, (John Wiley, New York, 2007) 3rd ed.
\end{thebibliography}
\end{document}